\documentclass[12pt,a4paper]{article} 
\usepackage{a4wide}

\title{Solitons and Other Extended Field Configurations}
 \author{R.\ S.\ Ward
 \bigskip
\\Department of Mathematical Sciences,  \\ University of
Durham, \\Durham DH1 3LE}
\newcommand{\mn}{\medskip\noindent}
\newcommand{\tr}{\mathop{\rm tr}\nolimits}
\newcommand{\sech}{\mathop{\rm sech}\nolimits}

\newcommand{\RR}{{\bf R}}
\newcommand{\pa}{\partial}
\newcommand{\ii}{{\rm i}}

\newcommand{\zb}{\overline{z}}
\newcommand{\Wb}{\overline{W}}

\begin{document}
\maketitle

\abstract{\noindent Article for the forthcoming Encyclopedia of
Mathematical Physics, to be published by Elsevier.  Covers kinks \&\
breathers, sigma-models \&\ Skyrmions, abelian-Higgs vortices,
monopoles, Yang-Mills instantons, and Q-balls.
}
\newpage

%%%%%%%%%%%%%%%%%%%%%%%%%%%%%%%%%%%%%%%%%%%%%%%%%%%%%%%%%

\section*{Introduction}

A soliton is a localized lump (or string or wall etc) of energy, which can
move without distortion, dispersion or dissipation, and which is stable under
perturbations (and collisions with other solitons).  The word was coined
by Zabusky and Kruskal in 1965, to describe a solitary wave with particle-like
properties (as in electron, proton etc).  Solitons are relevant
to numerous areas of physics --- condensed-matter, cosmology, fluids/plasmas,
biophysics (eg DNA), nuclear physics, high-energy physics etc. 
Mathematically, they are modelled as solutions of appropriate partial
differential equations.

Systems which admit solitons may be classified according to the mechanism
by which stability is ensured.  Such mechanisms include complete integrability,
nontrivial topology plus dynamical balancing, and Q-balls/breathers.

Sometimes the term `soliton' is used in a restricted sense, to refer
to stable localized lumps which have purely elastic interactions: solitons
which collide without any radiation being emitted.  This is possible only in
very special systems, namely those that are completely-integrable.
For these systems, soliton stability (and the elasticity of collisions) arise
from a number of characteristic properties, including:
a precise balance between dispersion and nonlinearity, solvability by the inverse
scattering transform from linear data, infinitely many conserved quantities,
a Lax formulation (associated linear problem), and B\"acklund transformations.
Examples of such integrable soliton systems are the sine-Gordon,
Korteweg-deVries and nonlinear Schr\"odinger equations.

The category of topological
solitons is the most varied, and includes such examples as kinks, vortices,
monopoles, Skyrmions and instantons. The requirement for these of `dynamical
balancing' can be understood in terms of Derrick's theorem, which provides
necessary conditions for a classical field theory to admit static localized
solutions.  The Derrick argument involves studying what happens to the energy
of a field when one changes the scale of space.  If one has a scalar field
(or multiplet of scalar fields) $\phi$, and/or a gauge field $F_{\mu\nu}$, then the
static energy $E$ is the sum of terms such as
\[
E_0=\int V(\phi)\,d^nx, \quad E_d=\int T_d(D_j\phi)\,d^nx, \quad
       E_F=\int F_{jk}F_{jk}\,d^nx,
\]
where each integral is over ($n$-dimensional) space ${\bf R}^n$, $D_j\phi$
denotes the covariant spatial derivative of $\phi$, and $T_d(\xi_j)$ is a
real-valued polynomial of degree $d$.  In particular, for example, we could
have $T_2(D_j\phi)=(D_j\phi)(D_j\phi)$: the standard gradient term.  Under
the dilation $x^j\mapsto\lambda x^j$, these functionals transform as
\[
E_0\mapsto\lambda^{-n}E_0, \quad E_d\mapsto\lambda^{d-n}E_d, \quad 
E_F\mapsto\lambda^{4-n}E_F.
\]
In order to have a static solution (critical point of the static energy
functional), one
needs to have a zero exponent on $\lambda$, and/or a balance between positive
and negative exponents.  A negative exponent indicates a compressing force
(tending to implode a localized lump), whereas a positive exponent indicates
an expanding force; so to have a static lump solution, these two forces have to
balance each other.  For $n=1$, a system involving only a scalar field,
with terms of the form $E_0$ and $E_2$,  can admit static solitons --- for
example kinks; the scaling argument implies a Virial Theorem, which in this
case says that $E_0=E_2$.  For $n=2$, one can have a scalar system with only
$E_2$, since in this case the relevant exponent is zero --- for example the
2-dimensional sigma-model. Another $n=2$ example is that of vortices in the
Abelian-Higgs model, where the energy contains terms $E_0$, $E_2$ and $E_F$.
For $n=3$, interesting systems have $E_2$ together
with either $E_4$ (for example Skyrmions) or $E_F$ (for example monopoles).
An $E_0$ term is optional in these cases; its presence affects, in particular,
the long-range properties of the solitons.  For $N=4$, one can have instantons
in a pure gauge theory (term $E_F$ only).

It should be noted that if there are no restrictions on the fields
$\phi$ and $A_j$ (such as arise, for example, from non-trivial topology),
then there is a more obvious mode of instability, which will inevitably be
present:  $\phi\mapsto\mu\phi$ and/or $A_j\mapsto\mu A_j$, where
$0\leq\mu\leq1$.  In other words, the fields can simply be scaled away
altogether, so that the height of the soliton (and its energy) go smoothly
to zero.  This can be prevented by non-trivial topology.

Another way of preventing solitons from shrinking is to allow the field to have
some `internal' time-dependence, so that it is stationary rather than static.
For example, one could allow the complex scalar field $\phi$ to have
the form $\phi = \psi\exp(\ii\omega t)$, where $\psi$ is independent of time $t$.
This leads to something like a centrifugal force, which can have a stabilizing
effect in the absence of Skyrme or magnetic terms.  The corresponding
solitons are $Q$-balls.

%%%%%%%%%%%%%%%%%%%%%%%%%%%%%%%%%%%%%%%%%%%%%%%%%%%%%%%%%%%

\section*{Kinks and Breathers}

The simplest topological solitons are kinks, in systems involving a
real-valued scalar field $\phi(x)$ in one spatial dimension.
The dynamics is governed by the Lagrangian density
\[
 {\cal L} = \frac{1}{2}\left[(\phi_t)^2-(\phi_x)^2-W(\phi)^2\right],
\]
where $W(\phi)$ is some (fixed) smooth function.  The system can admit
kinks if $W(\phi)$ has at least two zeros, for example $W(A)=W(B)=0$ with
$W(\phi)>0$ for $A<\phi<B$.  Two well-known systems are: sine-Gordon, where
$W(\phi)=2\sin(\phi/2)$, $A=0$, $B=2\pi$; and phi-four, where
$W(\phi)=1-\phi^2$, $A=-1$, $B=1$.  The corresponding field equations are
the Euler-Lagrange equations for ${\cal L}$; for example, the sine-Gordon
equation is
\begin{equation} \label{SG1}
  \phi_{tt} - \phi_{xx} + \sin\phi = 0.
\end{equation}

Configurations satisfying the boundary conditions $\phi\to A$ as
$x\to-\infty$ and $\phi\to B$ as $x\to\infty$ are called kinks (and
the corresponding ones with $x=\infty$ and $x=-\infty$ interchanged are
antikinks).  For kink (or antikink) configurations, there is a lower bound,
called the Bogomol'nyi bound, on the static energy $E[\phi]$; for kink
boundary conditions, we have
\begin{eqnarray*}
  E[\phi] &=& \frac{1}{2}
    \int_{-\infty}^{\infty} \bigl[ (\phi_x)^2 + W(\phi)^2 \bigr]\,dx\\
    &=&\frac{1}{2}\int_{-\infty}^{\infty} \bigl[\phi_x-W(\phi)\bigr]^2\,dx
        +\int_{-\infty}^{\infty}W(\phi) \phi_x \,dx\\
  &\geq& \int_A^B W(\phi)\,d\phi,
\end{eqnarray*}
with equality if and only if the Bogomol'nyi equation
\begin{equation} \label{Bog1}
  {d\phi \over dx} = W(\phi)
\end{equation}
is satisfied.  A static solution of the Bogomol'nyi equation is a kink
solution --- it is a static minimum of the energy functional in the kink
sector.  For example, for the sine-Gordon system we get
$E[\phi]\geq8$, with equality for the sine-Gordon kink
\[
  \phi(x) = 4 \tan^{-1}\exp(x-x_0);
\]
while for the phi-four system we get $E[\phi]\geq4/3$, with
equality for the phi-four kink
\[
  \phi(x) = \tanh(x-x_0).
\]

These kinks are stable topological solitons; the non-trivial topology
corresponds to the fact that the boundary value of $\phi(t,x)$ at $x=\infty$
is different from that at $x=-\infty$.  With trivial boundary conditions
(say $\phi\to A$ as $x\to\pm\infty$) stable static solitons are unlikely to
exist, but solitons with periodic time-dependence (which is this context
are called breathers) may exist.  For example, the sine-Gordon equation and
the non-linear Schr\"odinger equation both admit breathers --- but these
owe their existence to complete integrability.  By contrast, the phi-four
system (which is not integrable) does not admit breathers; a collision between
a phi-four kink and an antikink (with suitable impact speed) produces a
long-lived state which looks like a breather, but eventually decays into
radiation.

In lattice systems, however, breathers are more generic.  In a 1-dimensional
lattice system, the continuous space ${\bf R}$ is replaced by the lattice
${\bf Z}$, so $\phi(t,x)$ is replaced by $\phi_n(t)$ where $n\in{\bf Z}$.
The Lagrangian is
\[
 L = \frac{1}{2}\sum_n\left[(\dot\phi_n)^2-h^{-2}(\phi_{n+1}-\phi_n)^2
              -W(\phi_n)\right];
\]
here $h$ is a positive parameter, corresponding to the dimensionless ratio
between the lattice spacing and the size of a kink.  The continuum limit is
$h\to0$.  This system admits kink solutions as in the continuum case; and
for $h$ large enough, it admits breathers as well, but these disappear as
$h$ becomes small.

Interpreted in three dimensions, the kink becomes a domain wall separating two
regions in which the order parameter $\phi$ takes on distinct values; this has
applications in such diverse areas as cosmology and condensed-matter physics.

%%%%%%%%%%%%%%%%%%%%%%%%%%%%%%%%%%%%%%%%%%%%%%%%%%%%%%%

\section*{Sigma-Models and Skyrmions}

In a sigma-model or Skyrme system, the field is a map $\phi$ from
space-time to a Riemannian manifold $M$; usually $M$ is taken to be
a Lie group or a symmetric space.  The energy density of a static
field can be constructed as follows
(the Lorentz-invariant extension of this gives a relativistic Lagrangian
for fields on space-time).  Let $\phi^a$ be local coordinates on the
$m$-dimensional manifold $M$, let $h_{ab}$ denote the metric of $M$, and
let $x^j$ denote the spatial coordinates on space ${\bf R}^n$.
Define an $m\times m$ matrix $D$ by
\[
  {D_a}^b = (\partial_j \phi^c) h_{ac} (\partial_j \phi^b),
\]
where $\partial_j$ denotes derivatives with respect to the $x^j$.
Then the invariants ${\cal E}_2=\tr(D)=|\partial_j \phi^a|^2$ and
${\cal E}_4=\frac{1}{2}\left[(\tr D)^2-\tr(D^2)\right]$ can be
terms in the energy density, as well as a zeroth-order term
${\cal E}_0=V(\phi^a)$ not involving derivatives of $\phi$.
A term of the form ${\cal E}_4$ is called a Skyrme term.

The boundary condition on field configurations is that $\phi$ tends to
some constant value $\phi_0\in M$ as $|x|\to\infty$ in ${\bf R}^n$.
From the topological point of view, this compactifies ${\bf R}^n$ to
$S^n$.  In other words, $\phi$ extends to a map from $S^n$ to $M$;
and such maps are classified topologically by the homotopy group
$\pi_n(M)$.  For topological solitons to exist, this group has to be
non-trivial.

In one spatial dimension ($n=1$) with $M=S^1$ (say), the expression
${\cal E}_4$ is identically zero, and we just have kink-type systems
such as sine-Gordon.  The simplest two-dimensional example ($n=2$) is
the O(3) sigma model, which has $M=S^2$ with its standard metric.
In this system, the field is often expressed
as a unit 3-vector field $\vec\phi=(\phi^1,\phi^2,\phi^3)$, with
${\cal E}_2=(\partial_j\vec\phi)\cdot(\partial_j\vec\phi)$.
Here the configurations are classified topologically by their degree
(or winding number, or topological charge) $N\in\pi_2(S^2)\cong{\bf Z}$,
which equals
\[
N = \frac{1}{4\pi}
 \int\vec\phi\cdot\partial_1\vec\phi\times\partial_2\vec\phi\,dx^1\,dx^2.
\]
It is often convenient, instead of $\vec\phi$, to use a single complex-valued
function $W$ related to $\vec\phi$ by the stereographic projection
$W=(\phi^1+i\phi^2)/(1-\phi^3)$.  In terms of $W$, the formula for the degree
$N$ is
\[
 N = \frac{\ii}{2\pi}\int\frac{W_1\Wb_2-W_2\Wb_1}{(1+|W|^2)^2}
                             \,dx^1\,dx^2\,,
\]
and the static energy is (with $z=x^1 +\ii x^2$)
\begin{eqnarray*}
  E &=& \int{\cal E}_2\,d^2x\\
    &=& 8\int\frac{|W_z|^2 + |W_{\zb}|^2}{(1+|W|^2)^2}\,d^2x \\
    &=& 16\int\frac{|W_{\zb}|^2}{(1+|W|^2)^2}\,d^2x
         + 8\int\frac{|W_z|^2 - |W_{\zb}|^2}{(1+|W|^2)^2}\,d^2x \\
    &=&  16\int\frac{|W_{\zb}|^2}{(1+|W|^2)^2}\,d^2x + 8\pi N\,.
\end{eqnarray*}
From this, one sees that $E$ satisfies the Bogomol'nyi bound $E\geq8\pi N$;
and that minimal-energy solutions correspond to solutions of the Cauchy-Riemann
equations $W_{\zb}=0$.  To have finite energy, $W(z)$ has to be a rational
function, and so solutions with with winding number $N$
correspond to rational meromorphic functions $W(z)$, of degree $|N|$.
(If $N<0$, then $W$ is a rational function of $\zb$.)
The energy is scale-invariant (conformally-invariant), and consequently
these solutions
are not solitons --- they are not quite stable, since their size is not
fixed.  Adding terms ${\cal E}_4$ and ${\cal E}_0$ to the energy density
fixes the soliton size, and the resulting two-dimensional Skyrme systems
admit true topological solitons.

The three-dimensional case ($n=3$), with $M$ being a simple Lie group,
is the original Skyrme model of nuclear physics.  If $M=SU(2)$, then
the integer $N\in\pi_3(SU(2))\cong{\bf Z}$ is interpreted as the
baryon number.  The (quantum) excitations of the $\phi$-field correspond
to the pions, whereas the (semi-classical) solitons correspond to the
nucleons.  This model emerges as an effective theory of QCD, in the limit
where the number of colours is large.  If we express the field as a function
$U(x^j)$ taking values in a Lie group, then $L_j=U^{-1}\partial_j U$ takes
values in the corresponding Lie algebra, and ${\cal E}_2$ and ${\cal E}_4$
take the form
\begin{eqnarray*}
  {\cal E}_2 &=& -\frac{1}{2}\tr(L_j L_j) \\
  {\cal E}_4 &=& -\frac{1}{16}\tr\left([L_j, L_k][L_j, L_k]\right).
\end{eqnarray*}
The static energy density in the basic Skyrme system is the sum of these two
terms. The static energy satisfies a Bogomol'nyi bound $E\geq12\pi^2|N|$,
and it is believed that stable solitons (Skyrmions) exist for each value
of $N$.  Classical Skyrmions have been investigated numerically;
and, for values of $N$ up to about 25, they turn out to resemble polyhedral
shells.  Comparison with nucleon phenomenology requires
semi-classical quantization, and this leads to results which are at least
qualitatively correct.

A variant of the Skyrme model is the Skyrme-Faddeev system, which has
$n=3$ and $M=S^2$; the solitons in this case resemble loops which can be
linked or knotted, and which are classified by their Hopf number
$N\in\pi_3(S^2)$.  In this case, the energy satisfies a lower bound of
the form $E\geq cN^{3/4}$.  Numerical experiments indicate that
for each $N$, there is a minimal-energy solution with Hopf number $N$,
and with energy close to this topological lower bound.

%%%%%%%%%%%%%%%%%%%%%%%%%%%%%%%%%%%%%%%%%%%%%%%%%%%%%%%

\section*{Abelian-Higgs Vortices.}

Vortices live in two spatial dimensions, and viewed in three dimensions
are string-like;  two applications are as cosmic strings and as magnetic flux
tubes in superconductors.  They occur as static topological solitons in the
the abelian Higgs model (or Ginzberg-Landau model), and involve a magnetic
field $B=\pa_1 A_2 - \pa_2 A_1$, coupled to a complex scalar field $\phi$,
on the plane ${\bf R}^2$.  The energy density is
\begin{equation} \label{vortex-E}
   {\cal E} = \frac{1}{2}(D_j\phi)(\overline{D_j\phi}) + \frac{1}{2} B^2
             + \frac{1}{8}\lambda(1-|\phi|^2)^2\,,
\end{equation}
where $D_j\phi := \partial_j \phi - i A_j\phi$, and where $\lambda$ is a positive
constant.  The boundary conditions are
\begin{equation} \label{vortex-BCs}
  D_j\phi = 0, \quad B=0, \quad |\phi|=1
\end{equation}
as $r\to\infty$.  If we think of a very large circle $C$ on ${\bf R}^2$,
so that (\ref{vortex-BCs}) holds on $C$,  then $\phi\bigr|_C$ is a map from the
circle $C$ to the circle of unit radius in the complex plane, and therefore
it has an integer winding number $N$.  Thus configurations are labelled
by this vortex number~$N$.

Note that if ${\cal E}$ vanishes, then $B=0$ and $|\phi|=1$: the gauge
symmetry is spontaneously broken, and the photon `acquires a mass': this is
a standard example of spontaneous symmetry breaking.

The total magnetic flux $\int B\,d^2x$ equals $2\pi N$; a proof of this is
as follows.  Let $\theta$ be the usual polar coordinate around $C$.
Because $|\phi|=1$ on $C$, we can write $\phi=\exp[\ii f(\theta)]$ for
some function $f$; this $f$ need not be single-valued, but must satisfy
$f(2\pi)-f(0)=2\pi N$ with $N$ being an integer (in order that $\phi$ be
single-valued).  In fact, this defines the winding number.  Now
since  $D_j\phi = \pa_j\phi - \ii A_j\phi = 0$ on $C$, we have
\[
  A_j = -\ii\phi^{-1}\pa_j\phi = \pa_j f
\]
on $C$.  So, using Stokes' theorem, we get
\begin{eqnarray*}
\int_{\RR^2} B \,d^2x
       &= &\int_C A_j \, dx^j  \\
       &= &\int_0^{2\pi} \frac{df}{d\theta}\,d\theta \\
       &= &2\pi N\,.
\end{eqnarray*}

If $\lambda=1$, then the total energy
$E=\int{\cal E}\,d^2x$ satisfies the Bogomol'nyi bound $E \geq \pi N$;
and $E=\pi N$ if and only if a set of partial differential equations
(the Bogomol'nyi equations) are satisfied.
Since like charges repel, the magnetic force between
vortices is repulsive.  But there is also a force from the Higgs field,
and this is attractive.  The balance between the two forces is determined
by $\lambda$: if $\lambda>1$, then vortices repel each other;
while if $\lambda<1$, then vortices attract.  In the critical case $\lambda=1$,
the force between vortices is exactly balanced, and there
exist static multi-vortex solutions.  In fact, one has the following:
given $N$ points in the plane, there exists an $N$-vortex
solution of the Bogomol'nyi equations (and hence of the full field equations)
with $\phi$ vanishing at the chosen points (and nowhere else).
All static solutions are of this form.  These solutions cannot, however, be
written down explicitly in terms of elementary functions (except of course
for $N=0$).

%%%%%%%%%%%%%%%%%%%%%%%%%%%%%%%%%%%%%%%%%%%%%%%%%%%%%%%%%%%

\section*{Monopoles}

The abelian Higgs model does not admit 3-dimensional solitons, but a
non-abelian generalization does --- such non-abelian Higgs solitons are
called magnetic monopoles.  The field content,
in the simplest version, is as follows. First, there is a gauge
(Yang-Mills) field $F_{\mu\nu}$, with gauge potential $A_\mu$, and with
the gauge group being a simple Lie group $G$.  Secondly, there is a Higgs
scalar field $\phi$, transforming under the adjoint representation of $G$
(thus $\phi$ takes values in the Lie algebra of $G$).  For simplicity, take
$G$ to be SU(2) in what follows. So we may write $A_\mu=iA^a_\mu\sigma_a$,
$F_{\mu\nu}=iF^a_{\mu\nu}\sigma_a$ and $\phi=i\phi^a\sigma_a$, where $\sigma_a$
are the Pauli matrices.  The energy of static ($\partial_0\phi=0=\partial_0A_j$),
purely-magnetic ($A_0=0$) configurations is
\[
  E=\int\left[\frac{1}{2}B^a_j B^a_j + \frac{1}{2}(D_j\phi)^a(D_j\phi)^a
     +\frac{1}{4}\lambda(1-\phi^a\phi^a)^2\right]\,d^3x,
\]
where $B^a_j=\frac{1}{2}\epsilon_{jkl}F_{kl}$ is the magnetic field.
The boundary conditions are $B^a_j\to0$ and $\phi^a\phi^a\to1$ as $r\to\infty$;
so $\phi$ restricted to a large spatial 2-sphere becomes a map from $S^2$
to the unit 2-sphere in the Lie algebra $su(2)$, and as such it has a degree
$N\in{\bf Z}$.  An analytic expression for $N$ is
\begin{equation} \label{mon1}
  \int B^a_j (D_j\phi)^a \,d^3x = 2\pi N.
\end{equation}
At long range, the field resembles an isolated magnetic pole (a Dirac magnetic
monopole), with
magnetic charge $2\pi N$.  Asymptotically, the SU(2) gauge symmetry is
spontaneously broken to U(1), which is interpreted as the electromagnetic gauge
group.

In 1974, it was observed pointed out that this system
admits a smooth, finite-energy, stable, spherically-symmetric $N=1$ solution
--- this is the 't Hooft-Polyakov monopole.  There is a Bogomol'nyi lower
bound on the energy $E$: from $0\leq(B+D\phi)^2=B^2+(D\phi)^2+2B\cdot D\phi$
we get
\begin{equation} \label{mon2}
  E \geq 2\pi N + \int\frac{1}{4}\lambda(1-\phi^a\phi^a)^2\,d^3x,
\end{equation}
where use has been made of (\ref{mon1}). The inequality (\ref{mon2})
is saturated if and only if we go to the Prasad-Sommerfield limit $\lambda=0$,
and the Bogomol'nyi equations
\begin{equation} \label{mon3}
  (D_j\phi)^a = -B^a_j
\end{equation}
hold.  The corresponding solitons are called BPS monopoles.

The Bogomol'nyi equations (\ref{mon3}), together with the boundary conditions
described above, form a completely-integrable elliptic system of partial
differential equations.  For any positive integer $N$, the space of BPS
monopoles of charge $N$, with gauge freedom factored out, is parametrized by
a $(4N-1)$-dimensional manifold ${\cal M}_N$.  This is the moduli space of
$N$ monopoles.  Roughly speaking, each monopole has a position in space
(3 parameters) plus a phase (1 parameter), making a total of $4|N|$ parameters;
an overall phase can be removed by a gauge transformation, leaving
$4|N|-1$ parameters.  In fact, it is often useful to retain the overall phase,
and to work with the corresponding  $4|N|$-dimensional manifold
$\widetilde{{\cal M}}_N$.  This manifold has a natural metric, which corresponds
to the expression for the kinetic energy of the system.  A point in
$\widetilde{{\cal M}}_N$ represents an $N$-monopole configuration, and the
slow-motion dynamics of $N$ monopoles corresponds to geodesics on
$\widetilde{{\cal M}}_N$; this is the geodesic approximation of monopole dynamics.

The $N=1$ monopole is spherically-symmetric, and the corresponding fields
take a simple form; for example, the Higgs field of a 1-monopole located
at $r=0$ is
\[
  \phi^a=\left[\frac{\coth(2r)}{r} - \frac{1}{2r^2}\right]x^a.
\]
For $N>1$, the expressions tend to be less explicit; but monopole solutions
can nevertheless be characterized in a fairly complete way.
The Bogomol'nyi equations (\ref{mon3}) are a dimensional reduction of the
self-dual Yang-Mills equations in ${\bf R}^4$, and
BPS monopoles correspond to holomorphic vector bundles over a certain
two-dimensional complex manifold (`mini-twistor space').  This leads
to various other characterizations of monopole solutions, for example
in terms of certain curves (`spectral curves') on mini-twistor space,
and in terms of solutions of a set of ordinary differential equations
called the Nahm equations.  Having all these descriptions enables one to
deduce much about the monopole moduli space, and to characterize many
monopole solutions.
In particular, there are explicit solutions of the Nahm equations
involving elliptic functions, which correspond to monopoles with certain
discrete symmetries, such as a 3-monopole with tetrahedral symmetry,
and a 4-monopole with the appearance and symmetries of a cube.

%%%%%%%%%%%%%%%%%%%%%%%%%%%%%%%%%%%%%%%%%%%%%%%%%%%%%%%%%%%%%%%%

\section*{Yang-Mills Instantons}

Consider gauge fields in 4-dimensional Euclidean space ${\bf R}^4$,
with gauge group $G$. For simplicity in what follows, take $G$
to be SU(2); one can extend much of the structure to more general
groups, for example the simple Lie groups.  Let $A_\mu$ and
$F_{\mu\nu}$ denote the gauge potential and gauge field.
The Yang-Mills action is
\begin{equation}\label{YMaction}
 S=-\frac{1}{4} \int \tr\left(F_{\mu\nu}F_{\mu\nu}\right)\,d^4x,
\end{equation}
where we assume a boundary condition, at infinity in ${\bf R}^4$,
such that this integral converges.  The Euler-Lagrange equations
which describe critical points of the functional $S$ are
the Yang-Mills equations
\begin{equation}\label{YMeqn}
    D_{\mu} F_{\mu\nu} = 0.
\end{equation}
Finite-action Ynag-Mills fields are called instantons.
The Euclidean action (\ref{YMaction}) is used in the path-integral
approach to quantum gauge field theory, and therefore instantons
are crucial in understanding the path integral.

The dual of the field tensor $F_{\mu\nu}$ is
\[
 *F_{\mu\nu}=
    \frac{1}{2}\varepsilon_{\mu\nu\alpha\beta} F_{\alpha\beta}\,;
\]
the gauge field is self-dual if $*F_{\mu\nu}=F_{\mu\nu}$, and
anti-self-dual if $*F_{\mu\nu}=-F_{\mu\nu}$.  In view of the
Bianchi identity $D_{\mu} *F_{\mu\nu} = 0$, any self-dual or
anti-self-dual gauge field is automatically a solution of the
Yang-Mills equations (\ref{YMeqn}).  This fact also follows
from the discussion below, where we see that self-dual instantons
give local minima of the action.

The Yang-Mills action (and Yang-Mills equations) are conformally
invariant; and any finite-action solution of the Yang-Mills
equations on ${\rm R}^4$ extends smoothly to the conformal
compactification $S^4$. Gauge fields on $S^4$, with gauge group SU(2),
are classified topologically by an integer $N$, namely the second
Chern number
\begin{equation}\label{YMc2}
  N=c_2=-\frac{1}{8\pi^2}
     \int\tr\left(F_{\mu\nu}\,*F_{\mu\nu}\right)\,d^4x.
\end{equation}
From (\ref{YMaction}) and (\ref{YMc2}) we get a topological lower
bound on the action, as follows:
\begin{eqnarray*}
  0 &\leq& -\int\tr\left(F_{\mu\nu}-*F_{\mu\nu}\right)
          \left(F_{\mu\nu}-*F_{\mu\nu}\right)\,d^4x\\
    &=& 8S - 16\pi^2 N;
\end{eqnarray*}
and so $S\geq2\pi^2 N$, with equality if and only if the field is self-dual.
If $N<0$, we get $S\geq2\pi^2 |N|$, with equality if and only if $F$ is
anti-self-dual. So the self-dual (or anti-self-dual) fields minimize the
action in each topological class.

For the remainder of this section, we restrict to self-dual instantons
with instanton  number $N>0$. The space (moduli space) of such instantons,
with gauge-equivalence factored out, is a $(8N-3)$-dimensional real manifold.
In principle, all these gauge fields can be constructed using
algebraic-geometry (twistor) methods: instantons correspond to holomorphic
vector bundles over complex projective 3-space (twistor space).
 One large class of solutions which can be
written out explicitly is as follows; for $N=1$ and $N=2$ it gives all
instantons, while for $N\geq3$ it gives a $(5N+4)$-dimensional subfamily
of the full $(8N-3)$-dimensional solution space. The gauge potentials in
this class have the form
\begin{equation}\label{YMansatz}
   A_\mu = {\rm i} \sigma_{\mu\nu}\partial_\nu \log\phi,
\end{equation}
where the $\sigma_{\mu\nu}$ are constant matrices (antisymmetric in $\mu\nu$)
defined in terms of the Pauli matrices $\sigma_a$ by
\begin{eqnarray*}
  \sigma_{10} &=& \sigma_{23} = \frac{1}{2}\sigma_1, \\
  \sigma_{20} &=& \sigma_{31} = \frac{1}{2}\sigma_2, \\    
  \sigma_{30} &=& \sigma_{12} = \frac{1}{2}\sigma_3.
\end{eqnarray*}
The real-valued function $\phi=\phi(x^\mu)$ is a solution of the
4-dimensional Laplace equation given by
\[
 \phi(x^\mu)=
    \sum_{k=0}^N \frac{\lambda_k}{(x^\mu-x_k^\mu)(x^\mu-x_k^\mu)},
\]
where the $x_k^\mu$ are $N+1$ distinct points in ${\bf R}^4$, and the
$\lambda_k$ are $N+1$ positive constants: a total of $5N+5$ parameters.
It is clear from (\ref{YMansatz}) that the overall scale of $\phi$ is
irrelevant, leaving a $(5N+4)$-parameter family. For $N=1$ and $N=2$,
symmetries reduce the parameter count further, to 5 and 13 respectively.
Although $\phi$ has poles
at the points $x=x_k$, the gauge potentials are smooth (possibly after a
gauge transformation).

Finally, it is worth noting that (as one might expect) there is a gravitational
analogue of the gauge-theoretic structures described here.  In other words, one
has self-dual gravitational instantons --- these are 4-dimensional
Riemannian spaces for which the conformal-curvature tensor (the Weyl tensor)
is self-dual, and the Ricci tensor satisfies Einstein's
equations $R_{\mu\nu}=\Lambda g_{\mu\nu}$. As before, such spaces can be
constructed using a twistor-geometrical correspondence.

%%%%%%%%%%%%%%%%%%%%%%%%%%%%%%%%%%%%%%%%%%%%%%%%%%%%%%%%%%%%%%%%

\section*{Q-Balls}

A Q-ball (or nontopological soliton) is a soliton which has a periodic
time-dependence in a degree of freedom which corresponds to a global
symmetry.  The simplest class of Q-ball systems involves a complex
scalar field $\phi$, with an invariance under the constant phase
transformation $\phi\mapsto e^{i\theta}\phi$; the Q-balls are soliton
solutions of the form
\begin{equation}\label{QB1}
  \phi(t,{\bf x})=e^{i\omega t}\psi({\bf x}),
\end{equation}
where $\psi({\bf x})$ is a complex scalar field depending only on the
spatial variables {\bf x}.  The best-known case is the 1-soliton solution
\[
   \phi(t,x)=a\sqrt{2}\exp(ia^2t)\sech(ax)
\]
of the nonlinear Schr\"odinger equation $i\phi_t+\phi_{xx}+\phi|\phi|^2=0$.

More generally, consider a system (in $n$ spatial dimensions) with Lagrangian
\[
  {\cal L}=\frac{1}{2} (\partial_{\mu}\phi)(\partial^{\mu}\phi)-U(|\phi|),
\]
where $\phi(x^\mu)$ is a complex-valued field.  Associated with the global
phase symmetry is the conserved Noether charge
$Q = \int {\rm Im}(\bar\phi\phi_t)\,d^nx$.
Minimizing the energy of a configuration subject to $Q$ being fixed
implies that $\phi$ has the form (\ref{QB1}). Without loss of generality,
we may take $\omega\geq0$.  Note that $Q=\omega I$, where
$I = \int|\psi|^2\,d^nx$.  The energy of a configuration of the form (\ref{QB1})
is $E = E_q + E_k + E_p$, where
\begin{eqnarray*}
   E_q &=& \frac{1}{2}\int |\pa_j\psi|^2 \,d^nx, \cr
   E_k &=& \frac{1}{2} I\omega^2 = \frac{1}{2} Q^2/I, \cr
   E_p &=& \int U(|\psi|)\,d^nx.
\end{eqnarray*}
Let us take $U(0)=0=U'(0)$, with the field satisfying the boundary condition
$\psi\to0$ as $r\to\infty$.

A stationary Q-lump is a critical point of the energy functional $E[\psi]$,
subject to Q having some fixed value.  The usual (Derrick) scaling
argument shows that any stationary Q-lump must satisfy
\begin{equation}\label{QBscaling}
   (2-n)E_q - n E_p + n E_k = 0.
\end{equation}
For simplicity in what follows, let us take $n\geq3$.
Define $m>0$ by $U''(0)=m^2$; then, near spatial infinity, the
Euler-Lagrange equations give $\nabla^2\psi-(m^2-\omega^2)\psi=0$.
So in order to satisfy the
boundary condition $\psi\to0$ as $r\to\infty$, we need $\omega< m$.

It is clear from (\ref{QBscaling}) that if $U\geq\frac{1}{2}m^2|\psi|^2$
everywhere, then there can be no solution.  So
$K = {\rm min}\left[2U(|\psi|)/|\psi|^2\right]$ has to satisfy $K < m^2$.
Also, we have
\begin{equation}\label{omega-inequality}
 E_p = \int U \geq \frac{1}{2}KI = (K/\omega^2)E_k > (K/\omega^2)E_p,
\end{equation}
where the final inequality comes from (\ref{QBscaling}).
As a consequence, we see that $\omega^2$ is restricted to the range
\begin{equation}\label{omega}
   K < \omega^2 < m^2.
\end{equation}
An example which has been studied in some detail is $U(f)=f^2[1+(1-f^2)^2]$;
here $m^2=4$ and $K=2$, so the range of frequency for Q-balls in this system is
$\sqrt{2} < \omega < 2$.  The dynamics of Q-balls
in systems such as these turns out to be quite complicated.

%%%%%%%%%%%%%%%%%%%%%%%%%%%%%%%%%%%%%%%%%%%%%%%%%%%%%%%%%

\section*{Further Reading}

\mn M F Atiyah and N J Hitchin, The Geometry and Dynamics of Magnetic
    Monopoles (Princeton University Press, Princeton, 1988)

\mn S Coleman, Aspects of Symmetry (Cambridge University Press, Cambridge)

\mn P G Drazin and R S Johnson, Solitons: an Introduction
  (Cambridge University Press, Cambridge, 1989)

\mn P Goddard and P Mansfield, Topological structures in field theories.
   Reports on Progress in Physics 49 (1986), 725--781.

\mn A Jaffe and C Taubes, Vortices and Monopoles (Birkh\"auser, Boston, 1980)

\mn T D Lee and Y Pang, Nontopological Solitons.
   Physics Reports 221 (1992), 251--350.

\mn V G Makhankov, Y P Rubakov and V I Sanyuk, The Skyrme Model:
  Fundamentals, Methods, Applications (Springer, 1993)

\mn N S Manton and P M Sutcliffe, Topological Solitons
         (Cambridge University Press, Cambridge, 2004)

\mn R Rajaraman, Soliton and Instantons (North-Holland, 1982)

\mn C Rebbi and G Soliani, Solitons and Particles
    (World Scientific, Singapore, 1984)

\mn A Vilenkin and E P S Shellard, Cosmic Strings and Other
    Cosmological Defects (Cambridge University Press, Cambridge, 1994)

\mn R S Ward and R O Wells jr, Twistor Geometry and Field Theory
  (Cambridge University Press, Cambridge, 1990)

\mn W J Zakrzewski, Low Dimensional Sigma Models (IOP, Bristol, 1989)

%%%%%%%%%%%%%%%%%%%%%%%%%%%%%%%%%%%%%%%%%%%%%%%%%%%%%

\bigskip\bigskip
%\newpage

\noindent{\bf See also}

\noindent Topological defects and their homotopy classification

\noindent Instantons in gauge theory

\noindent 'tHooft-Polyakov monopoles

\noindent Vortices

%%%%%%%%%%%%%%%%%%%%%%%%%%%%%%%%%%%%%%%%%%

\bigskip\bigskip\noindent{\bf Keywords:} soliton, topological, kink,
  skyrmion, gauge, monopole, vortex, instanton, Q-ball, sigma-model,
  Bogomolnyi, moduli.

\end{document}